\PassOptionsToPackage{table}{xcolor}

\documentclass[sigconf]{acmart}
\AtBeginDocument{%
  }

\usepackage{amssymb}
\usepackage[table]{xcolor}
\usepackage{tcolorbox}

\usepackage{soul}
\usepackage{caption}
\usepackage{subcaption}

\usepackage{pifont}
\usepackage{multirow}

\newcommand{\highlight}[2]{\sethlcolor{#1}\hl{#2}}

\newcommand{\cmark}{\ding{51}}
\newcommand{\xmark}{\ding{55}}

\copyrightyear{2025}
\acmYear{2025}
\setcopyright{cc}
\setcctype{by}
\acmConference[RecSys '25]{Proceedings of the Nineteenth ACM Conference on
Recommender Systems}{September 22--26, 2025}{Prague, Czech Republic}
\acmBooktitle{Proceedings of the Nineteenth ACM Conference on Recommender
Systems (RecSys '25), September 22--26, 2025, Prague, Czech Republic}
\acmDOI{10.1145/3705328.3748013}
\acmISBN{979-8-4007-1364-4/2025/09}

\begin{document}

\title{Not Just What, But When: Integrating Irregular Intervals \\ to LLM for Sequential Recommendation}

\author{Wei-Wei Du}
\affiliation{
  \institution{Sony Group Corporation}
  \city{Tokyo}
  \country{Japan}
}
\email{weiwei.du@sony.com}
\orcid{0000-0002-0627-0314}

\author{Takuma Udagawa}
\affiliation{
  \institution{Sony Group Corporation}
  \city{Tokyo}
  \country{Japan}
}
\email{takuma.udagawa@sony.com}
\orcid{0009-0001-7441-5773}

\author{Kei Tateno}
\affiliation{
  \institution{Sony Group Corporation}
  \city{Tokyo}
  \country{Japan}
}
\email{kei.tateno@sony.com}
\orcid{0009-0000-8249-2659}

\renewcommand{\shortauthors}{Du et al.}

\begin{abstract}
Time intervals between purchasing items are a crucial factor in sequential recommendation tasks, whereas existing approaches focus on item sequences and often overlook by assuming the intervals between items are static.
However, dynamic intervals serve as a dimension that describes user profiling on not only the history within a user but also different users with the same item history.
In this work, we propose \textbf{IntervalLLM}, a novel framework that integrates interval information into LLM and incorporates the novel interval-infused attention to jointly consider information of items and intervals.
Furthermore, unlike prior studies that address the cold-start scenario only from the perspectives of users and items, we introduce a new viewpoint: \textit{the interval perspective} to serve as an additional metric for evaluating recommendation methods on the warm and cold scenarios.
Extensive experiments on 3 benchmarks with both traditional- and LLM-based baselines demonstrate that our IntervalLLM achieves not only 4.4\% improvements in average but also the best-performing warm and cold scenarios across all users, items, and the proposed interval perspectives.
In addition, we observe that the cold scenario from the interval perspective experiences the most significant performance drop among all recommendation methods.
This finding underscores the necessity of further research on interval-based cold challenges and our integration of interval information in the realm of sequential recommendation tasks.
Our code is available here: https://github.com/sony/ds-research-code/tree/master/recsys25-IntervalLLM.
\end{abstract}







\maketitle

\section{Introduction}
In sequential recommendation, next-item prediction aims to forecast the next item a user is likely to select based on the past behavior.
Recent state-of-the-art models have increasingly adopted LLM-based approaches \cite{liao2024llara, DBLP:journals/corr/abs-2306-05817}, which not only enhance textual understanding but also leverage reasoning ability and world knowledge.
Prior to the emergence of LLMs, several sequential recommendation models \cite{DBLP:conf/wsdm/LiWM20, DBLP:conf/aaai/DangYGJ0XSL23, DBLP:journals/corr/abs-2204-10851} incorporated interval information to capture the temporal dynamics of user behavior. 
With the advent of powerful LLMs, some studies \cite{DBLP:conf/iclr/0005WMCZSCLLPW24, DBLP:conf/nips/ZhouNW0023, 10.1145/3719207} integrated timestamps into LLM-based methods for time-series tasks, but none of these studies specifically focus on sequential recommendation.

\begin{figure}
    \vspace{10pt}
    \centering
    \includegraphics[width=\linewidth]{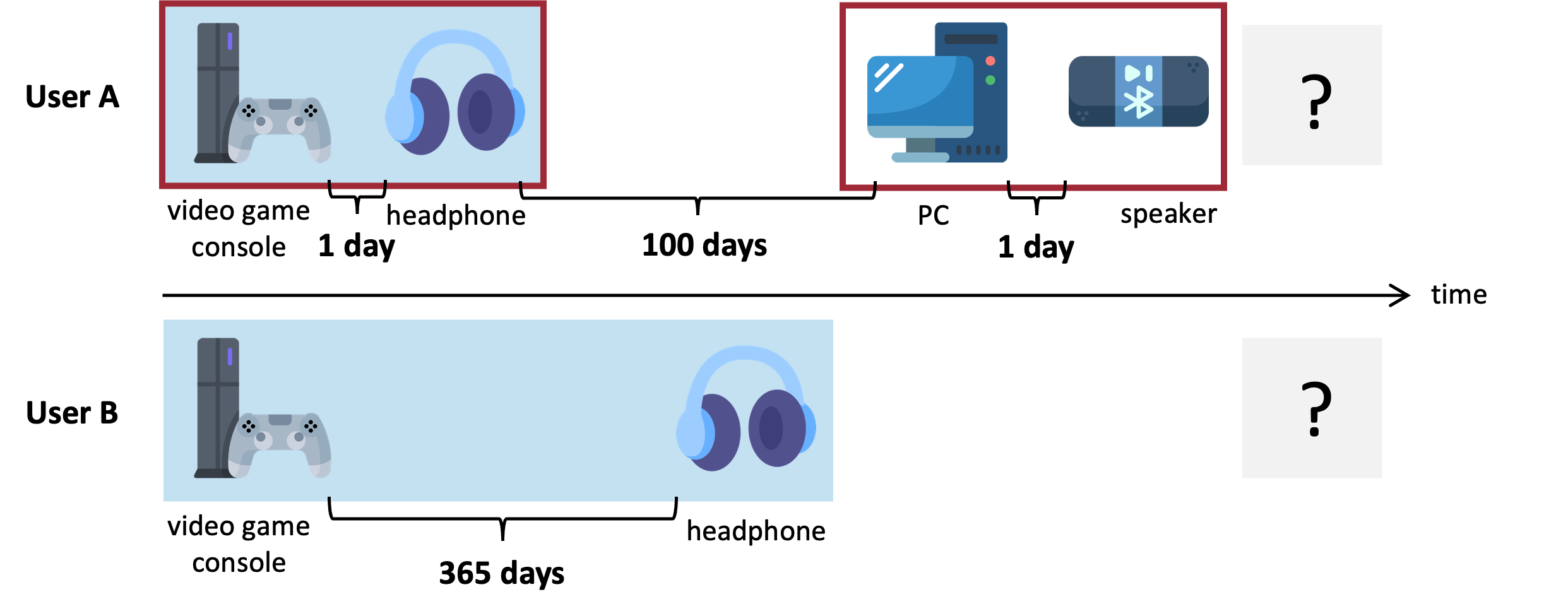}
    \caption{Examples of purchase histories for two users.}
    \label{fig:hypo}
\end{figure}

As shown in Figure \ref{fig:hypo}, existing models often consider the purchase item history as a sequence for sequential recommendation.
They neglect the influence of temporal intervals in shaping user behavior, as depicted by the red squares.
Even with identical item sequences, varying intervals (highlighted in blue for User A and User B) reveal differing item influences.
In this work, we take the first step toward integrating interval information into LLM-based recommender systems, bridging the gap between sequential recommendation and the interval modeling capabilities of LLMs.

\begin{figure*}
    \centering
    \includegraphics[width=\linewidth]{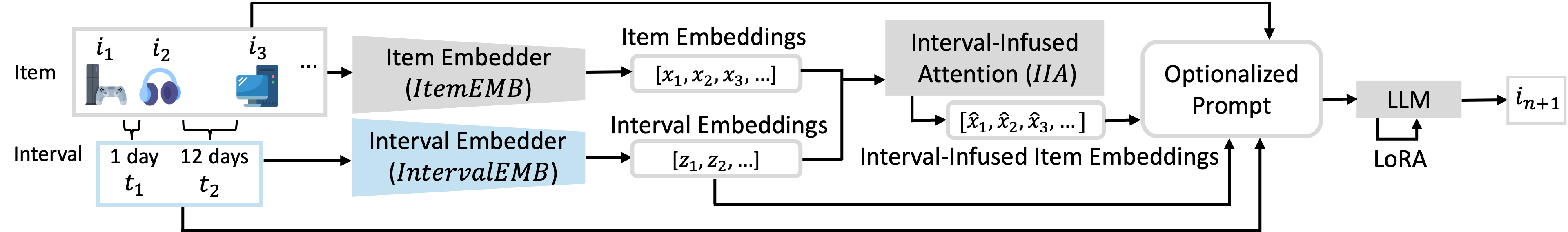}
    \caption{The proposed IntervalLLM. All LLM parameters except for those in LoRA are frozen.}
    \label{fig:framework}
\end{figure*}

Furthermore, as the number of users and items increases, user-item interaction data becomes increasingly sparse, making it challenging to model user and item behaviors with limited interactions.
This issue, commonly referred to as the cold-start problem, is frequently discussed in recommendation system research \cite{DBLP:journals/access/YuanH23, DBLP:journals/corr/abs-2501-01945}.
However, in real-world recommendation systems, where interaction data accumulate over long periods, the cold-start problem from a time-interval perspective, which refers to users whose interactions are characterized by relatively long time gaps, is another critical yet often overlooked issue.
User preferences evolve over time, and item trends change dynamically.
The temporal aspect of item sequences plays a significant role in distinguishing recent purchasing behaviors from historical ones.
To address this gap, we aim to benchmark existing sequential recommendation methods not only for cold-start users and items but also from an interval cold-start perspective.
Furthermore, we explore how leveraging LLM knowledge and interval-based information can enhance performance in scenarios where traditional methods struggle.

Our contributions are summarized as follows:
\begin{itemize}
   \item To the best of our knowledge, IntervalLLM is the first work to integrate time intervals into LLMs for sequential recommendation, enabling the model to capture user behavior based on the intervals between consecutive items.
   \item Beyond separately incorporating item and interval embeddings, we propose Interval-Infused Attention to capture the temporal relevance between items and intervals.
   \item We introduce a novel perspective on the cold-start problem by considering time intervals and reveal that existing methods suffer from significant performance drops in interval cold-start scenarios. Experiments demonstrate that IntervalLLM achieves the best performance in overall, warm, and even cold-start scenarios.
\end{itemize}

\section{Preliminaries}
\subsection{Related Work}
Recent advancements in LLMs have demonstrated their remarkable effectiveness due to their extensive world knowledge and strong generalization capabilities.
In the context of sequential recommendation tasks, two primary strategies have emerged for leveraging LLMs.
The first treats LLMs as feature extractors, using their embeddings to initialize existing recommendation models \cite{DBLP:journals/corr/abs-2104-07413, DBLP:conf/recsys/HarteZLKJF23}.
The second involves directly fine-tuning LLMs as recommenders, leveraging their vast pre-trained knowledge and advanced reasoning abilities \cite{DBLP:conf/recsys/HarteZLKJF23, DBLP:journals/corr/abs-2305-07622, DBLP:journals/corr/abs-2308-08434, DBLP:conf/recsys/BaoZZWF023}.
To further enrich recommendation representations, models such as MoRec \cite{yuan2023go}, A-LLMRec \cite{kim2024large}, and LLM-SRec \cite{kim2025lostsequencelargelanguage} incorporate user embeddings extracted from pre-trained collaborative filtering models.
However, these approaches exclude timestamp or interval information, thereby neglecting the temporal dimension inherent in sequential data.
To address this gap, we introduce a novel cold-start benchmark from the interval perspective and investigate effective strategies for integrating interval information into LLMs to enhance sequential recommendation performance.

\subsection{Sequential Recommendation Task}
In the context of sequential recommendation, a user interacts chronologically with a sequence of n item names [$i_1$, $i_2$, \dots, $i_n$] and corresponding intervals [$t_1$, $t_2$, \dots, $t_{n-1}$], where each $t_k$ represents the time difference between consecutive items.
To align with the generative capabilities of LLMs, a set of candidate items [$c_1$, $c_2$, \dots, $c_j$] is randomly sampled from the item pool for each user.
The goal of the sequential recommendation task can be formulated as:
\begin{equation}
    \begin{split}
    i_{n+1} = f([i_1, i_2, \dots, i_{n}], [t_1, t_2, \dots, t_{n-1}], [c_1, c_2, \dots, c_j])
    \end{split}
\end{equation}
where $i_{n+1}$ denotes the next item to be recommended at the $(n+1)$-th timestamp, and $f$ represents the recommendation algorithm.
Notably, the number of items in the sequence exceeds the number of time intervals by one, as each $t_k$ captures the temporal gap between consecutive items $i_k$ and $i_{k+1}$.

\section{Methodology}

\begin{figure*}
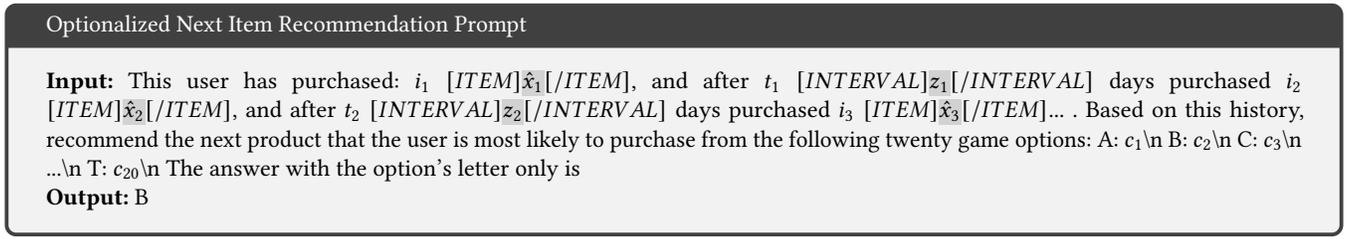

    \centering
    \begin{tcolorbox}[title=Optionalized Next Item Recommendation Prompt]
\textbf{Input:} This user has purchased: $i_1$ $[ITEM]$\highlight{gray!35}{$\hat{x}_1$}$[/ITEM]$, and after $t_1$ $[INTERVAL]$\highlight{gray!35}{$z_1$}$[/INTERVAL]$ days purchased $i_2$ $[ITEM]$\highlight{gray!35}{$\hat{x}_2$}$[/ITEM]$, and after $t_2$ $[INTERVAL]$\highlight{gray!35}{$z_2$}$[/INTERVAL]$ days purchased $i_3$ $[ITEM]$\highlight{gray!35}{$\hat{x}_3$}$[/ITEM]$... . Based on this history, recommend the next product that the user is most likely to purchase from the following twenty game options: A: $c_1$\textbackslash n B: $c_2$\textbackslash n C: $c_3$\textbackslash n ...\textbackslash n T: $c_{20}$\textbackslash n The answer with the option's letter only is 

\textbf{Output:} B
    \end{tcolorbox}
    \caption{The optionalized prompt template for the next item recommendation task. $i$ represents an item, $t$ denotes the corresponding interval, $c$ is the candidate item, \highlight{gray!35}{$\hat{x}$} is the attention weight, and \highlight{gray!35}{$z$} is the interval embedding. The special tokens $[ITEM]$ and $[/ITEM]$ mark the location of attention which is used for item embedding, while $[INTERVAL]$ and $[/INTERVAL]$ specify the position of the interval embedding.}
    \label{fig:prompt-template}
\end{figure*}

The framework of IntervalLLM is illustrated in Figure \ref{fig:framework}, which first maps items and intervals into a shared latent space, then employs an interval-infused attention mechanism to model their interactions.
Finally, we propose a novel optionalized prompt and leverage the LLM as the recommender.

\subsection{Behavior Representation Learning}
\textbf{Item Embedder.}
To apply LLMs as recommenders, following the previous work \cite{liao2024llara}, an item embedder is required to map each item into the language space, transforming the textual natural language of item names into embeddings.
Each item name associated with a user is converted into an embedding $x$, denoted as follows:
\begin{equation}
    x_{k} = ItemEMB(i_k),
\end{equation}
where $ItemEMB()$ represents the tokenizer and embedding layer.
Here, $k$ denotes the $k$-th item in the sequence.

\textbf{Interval Embedder.}
Previous work \cite{DBLP:journals/corr/abs-2310-07820} has proposed some regularization methods for processing numerical inputs.
In sequential recommendation, the scale of the interval is particularly crucial.
Therefore, we propose an interval embedder to encode the temporal scale $z$ as follows:
\begin{equation}
    z_{k-1} = IntervalEMB(t_{k-1}),
\end{equation}
where $IntervalEMB()$ denotes the embedding layer for the interval, which we implement as a simple MLP layer in our experiments.

\textbf{Interval-Infused Attention (IIA).}
To jointly model the relations between interval and items, an intuitive method is to directly add interval embeddings into the embedding sequence produced by the LLM. 
However, this fails to effectively integrate interval information into item embeddings. 
Therefore, we applied scaled dot-product attention \cite{DBLP:journals/corr/VaswaniSPUJGKP17} to separately characterize the importance of both items and intervals and then aggregate corresponding weights as interval-infused item embeddings.
Given a user as an example, we represent the item sequence with item textual embeddings $X=(x_1, \dots, x_{n})$ and the corresponding interval embeddings $Z=(0, z_1, \dots, z_{n-1})$.
To ensure alignment between the two sequences, a zero vector is padded at the beginning of the interval embeddings, making the lengths of $X$ and $Z$ identical.
The formula of the interval-infused attention $IIA()$ is derived as follows:
\begin{equation}
    Q_{z} = ZW^{Q_z}, K_{x} = XW^{K_x}, V_{x} = XW^{V_x},
\end{equation}
\begin{equation}
    IIA(X, Z) = softmax(\frac{Q_zK_x^T}{\sqrt{d_q}}+M)V_x,
\end{equation}
where $Q_{z}$ is the query matrix of $Z$, $K_{x}$ and $V_{x}$ denote the key and value matrices of $X$, and $M$ denotes the casual attention matrix to prevent information leakage from future tokens.
The projection matrices $W^{Q_z}, W^{K_x}, W^{V_x} \in \mathbb{R}^{d_{\text{llm}}\times d_q}$ are learnable parameters.

To consider user behaviors from different perspectives, we extend interval-infused attention with $h$ heads to learn interval-infused item embeddings:
\begin{equation}
    [\hat{x}_1, \dots, \hat{x}_n] = Concat(IIA_1, \dots, IIA_h)W^O,
\end{equation}
where $W^O\in \mathbb{R}^{hn\cdot d_{\text{llm}}\times n\cdot d_{\text{llm}}}$ is a linear layer.

\textbf{Optionalized Prompt.}
To structure the full sequence in a format compatible with LLM, we design a text prompt that effectively conveys the relevant information for LLM instruction tuning.
Most existing works consider sequential recommendation as a generation task, evaluating whether the ground truth is included in the generated sequence \cite{liao2024llara, DBLP:conf/nips/Liu0W000024}.
However, since the ground truth is often an item name that may be a full sentence rather than a single word, evaluating the generation task becomes more challenging.
For example, there are cases where the ground truth appears within the generated output but is accompanied by other item names, leading to ambiguity.
Additionally, the ground truth may be present in the generated output but with spelling errors or incorrect details, further complicating evaluation.
To address these challenges, we propose an option-based generation task, where alphabetic options (e.g., A, B, etc.) are added before each candidate item name as shown in Figure \ref{fig:prompt-template}.
This optional problem setting significantly reduces ambiguity and enables a more reliable evaluation of model performance.
In summary, the prompt includes task definition, user behavior sequence incorporating both items and intervals, and candidate options.
To align interval embeddings and interval-infused item embeddings with LLM (i.e., $d_\text{llm}$), all textual components are first processed through the initial layer of the LLM and then concatenated with embeddings.

\textbf{Instruction Fine-tuning by LoRA.}
To mitigate the gap between the capabilities of LLMs in general text generation and sequential recommendation tasks, LoRA \cite{DBLP:conf/iclr/HuSWALWWC22} is applied to fine-tune LLMs for sequential recommendation tasks.
The loss function of IntervalLLM is formulated as:
\begin{equation}
    L((i_{1..n}, t_{1..n-1}), i_{n+1}) = -log (P_\theta(i_{n+1}|(i_{1..n}, t_{1..n-1}))),
\end{equation}
where $\theta$ includes the parameters of the item embedder, the interval embedder, the interval-infused attention and the LoRA within the LLM's parameters.
\section{Experiments}
In this section, we conduct comprehensive experiments to answer the following key research questions: 
\begin{itemize}
  \item RQ1: How does IntervalLLM performs in three sequential recommendation datasets?
  \item RQ2: How do different components of IntervalLLM affect model performance?
  \item RQ3: How does IntervalLLM perform in the cold scenario from various perspectives?
\end{itemize}

\subsection{Experimental Setup}
\noindent \textbf{Datasets.}
We conduct experiments on three Amazon Reviews datasets \cite{hou2024bridging}: Video Games, CDs and Vinyl, and Books.
The characteristics of each dataset are summarized in Table \ref{tab:datasets}.
Following prior work \cite{DBLP:journals/corr/abs-1808-09781}, we use five-core datasets in which both users and items have at least five interactions each.

\begin{table}
    \caption{Dataset characteristics. Density is defined as the number of interactions per user (\#Interactions/\#Users).}
    \begin{tabular}{c|rrr|r}
    Dataset & \#User & \#Item & \#Interaction & Density\\
    \toprule
    Video Games & 94,762 & 25,612 & 814,586 & 8.59 \\
    CDs and Vinyl  & 123,876  & 89,370 &  1,552,764 & 12.53\\
    Books  & 776,370 & 495,063 & 9,488,297 & 12.22\\
\end{tabular}

    \label{tab:datasets}
\end{table}

\begin{table*}
    \caption{Overall performance with Hit Rate@1 ($\uparrow$) on three datasets. The best result in each column is in boldface, while the second-best result is underlined. * means that some of the predictions are not valid (e.g., misspelling).}

\begin{tabular}{c|c|@{\hskip 5pt}c@{\hskip 10pt}c@{\hskip 10pt}c@{\hskip 10pt}}
    Category & Method & Video Games & CDs and Vinyl & Books \\
    \toprule
    \multirow{2}{*}{Traditional} & GRU4Rec & 49.0\% & 46.5\% & 35.7\% \\
     & SASRec & 50.8\% & 50.9\% & 38.0\% \\
    \midrule
    Traditional + Interval & TiSASRec & 52.9\% & \underline{54.1\%} & 58.6\% \\
    \midrule
    \multirow{2}{*}{LLM-based} & LLaMA & 56.0\% & 45.2\%* & 60.2\% \\
     & LLaRA & 50.5\% & 49.6\%* & 60.0\%* \\
    \bottomrule
    \multirow{2}{*}{LLM-based + Interval} & LLaMA + Interval & \underline{56.3\%} & 48.7\% & \underline{61.1\%} \\
     & \cellcolor{gray!20}IntervalLLM (Ours) & \cellcolor{gray!20}\textbf{61.7\%}  & \cellcolor{gray!20}\textbf{55.4\%} &  \cellcolor{gray!20}\textbf{61.9\%}
\end{tabular}
    \label{tab:overall_performance}
\end{table*}

\noindent \textbf{Implementation Details.}
All LLM-based methods are trained for up to 5 epochs with a batch size of 64 and 20 candidates \cite{liao2024llara}, using LLaMA-2 (7B) as the backbone with an optionalized prompt.
For IntervalLLM, the LLM input dimension ($d_{llm}$) is 4096, embedding dimension ($d_{q}$) is 256, and the number of attention heads ($h$) is 2.

\noindent \textbf{Evaluation Metric.}
We follow previous work \cite{liao2024llara} to use Hit Ratio@1.
Notably, \cite{liao2024llara} further adds the validation ratio as an additional metric, as certain models may generate invalid responses, such as word mismatches with the candidate set; however, in our method, the validation ratio is consistently 100\% attributed to our optionalized prompt design.
We adopt a leave-one-out evaluation strategy following previous work \cite{DBLP:journals/corr/abs-1808-09781}, where the last item in each user's sequence is reserved for testing, the second-to-last item for validation, and the remaining items are used for training.

\noindent \textbf{Baselines.}
We compare four groups of models as our baselines: 
\begin{itemize}
   \item Traditional methods: GRU4Rec \cite{hidasi2016} and SASRec \cite{DBLP:journals/corr/abs-1808-09781}.
   \item Traditional method with interval: TiSASRec \cite{DBLP:conf/wsdm/LiWM20}.
   \item LLM-based methods: LLaMA \cite{DBLP:journals/corr/abs-2307-09288} and LLaRA \cite{liao2024llara}.
   \item LLM-based method with interval: LLaMA + Interval. As no existing method incorporates interval information into LLMs, we build a naive baseline by directly adding interval information into the prompt for LLaMA.
\end{itemize}

\begin{table}
    \caption{Ablation study of IntervalLLM on the Video Games dataset. $\triangle$ indicates applying timestamp as the text prompt. $IntervalEMB$ indicates adding the interval embedding. $IIA$ indicates adding the interval-infused embeddings.}
    \begin{tabular}{cccc|c}
    LLaMA & Interval & $IntervalEMB$ & $IIA$ & Hit Rate@1 ($\uparrow$) \\
    \toprule
    \cmark & \xmark & \xmark & \xmark & 56.0\% \\
    \midrule
    \cmark & $\triangle$ & \xmark & \xmark & 54.2\% \\
    \cmark & \cmark & \xmark & \xmark & 56.3\% \\
    \midrule
    \cmark & \cmark & \cmark & \xmark & 56.8\% \\
    \bottomrule
    \cmark & \cmark & \cmark & \cmark & \textbf{61.7\%} \\
\end{tabular}


    \label{tab:ablation}
\end{table}

\subsection{RQ1: Overall Performance Comparison}
Table \ref{tab:overall_performance} shows the recommendation performance of IntervalLLM compared with traditional sequential recommendation models and LLM approaches across three datasets, which draws the following observations: (1) LLM-based methods outperform traditional models, demonstrating that leveraging pre-trained LLM knowledge can enhance recommendation performance. 
This improvement is particularly notable in Books, where item names often carry rich semantic meaning, leading to substantial performance gains.
(2) LLaMA + Interval surpasses LLaMA across all three datasets, showcasing the effectiveness of adding interval information.
(3) IntervalLLM consistently achieves superior performance compared to all other methods on all datasets, highlighting not only the importance of incorporating interval information into LLMs but also the efficacy of the proposed interval-infused attention.
Overall, these findings emphasize the effectiveness of utilizing LLMs to capture the semantic meaning of items, incorporating interval information, designing attention to leverage item and interval, and employing optionalized prompts tailored for the recommendation task.

\subsection{RQ2: Ablation Study}
An extensive ablation study is conducted to validate the design of IntervalLLM, as presented in Table \ref{tab:ablation}. 
The comparison between the first row and second as well as third rows reveals that directly representing timestamp as the text hinders model performance, while using interval text helps understand temporal relations between items.
Furthermore, adding the interval embeddings from the interval embedder enables the model to learn the scale of intervals more effectively (row 4).
The boosted performance in row 5 indicates applying interval-infused attention enhances the model's ability to capture relations between items and their corresponding intervals. 
In summary, these results suggest that all proposed components contribute positively to the overall performance.

\begin{table*}
    \caption{Performance on Warm/Cold scenarios evaluated by Hit Rate@1 ($\uparrow$) on the Video Games dataset. \emph{Warm} represents the warm scenario, \emph{Cold} represents the cold-start scenario, and \emph{Diff.} = ((\emph{Cold} - \emph{Warm})/\emph{Warm}).} 
    \begin{tabular}{c|ccc|ccc|ccc||c}
    & \multicolumn{3}{c|}{User Perspective} & \multicolumn{3}{c|}{Item Perspective} & \multicolumn{3}{c||}{Interval Perspective} &  \\
    Method & \emph{Warm ($\uparrow$)} & \emph{Cold ($\uparrow$)} & \emph{Diff.} & \emph{Warm ($\uparrow$)} & \emph{Cold ($\uparrow$)} & \emph{Diff.} & \emph{Warm ($\uparrow$)} & \emph{Cold ($\uparrow$)} & \emph{Diff.} & Overall ($\uparrow$) \\
    \toprule
    GRU4Rec & 49.3\% & 48.8\% & \cellcolor{gray!20}-1.0\% & 53.3\% & 45.1\% & \cellcolor{gray!20}-15.4\% & 55.2\% & 43.7\% & \cellcolor{gray!20}-20.8\% & 49.0\% \\
    SASRec & 52.2\% & 50.1\% & \cellcolor{gray!20}-4.0\% & 54.7\% & 47.4\% & \cellcolor{gray!20}-13.3\% & 57.4\% & 45.2\% & \cellcolor{gray!20}-21.2\% & 50.8\% \\
    \midrule
    TiSASRec & 55.0\% & 52.3\% & \cellcolor{gray!20}-4.9\% & 54.3\% & 51.0\% & \cellcolor{gray!20}-6.1\% & 54.6\% & 49.1\% & \cellcolor{gray!20}-10.1\% & 52.9\% \\
    \midrule
    LLaMA & \underline{56.5\%} & 55.8\% & \cellcolor{gray!20}-1.2\% & 58.8\% & 53.7\% & \cellcolor{gray!20}-8.7\% & \underline{61.1\%} & 51.8\% & \cellcolor{gray!20}-15.2\% & 56.0\% \\
    LLaRA & 51.6\% & 50.0\% & \cellcolor{gray!20}-3.1\% & 54.0\% & 46.9\% & \cellcolor{gray!20}-13.1\% & 60.3\% & 43.3\% & \cellcolor{gray!20}-28.0\% & 50.5\% \\
    \bottomrule
    LLaMA + Interval & 56.1\% & \underline{56.2\%} & \cellcolor{gray!20}+0.2\% & \underline{59.1\%} & \underline{53.8\%} & \cellcolor{gray!20}-9.0\% & 60.6\% & \underline{53.3\%} & \cellcolor{gray!20}-12.0\% & \underline{56.3\%} \\
    IntervalLLM (Ours) & \textbf{61.6\%} & \textbf{61.8\%} & \cellcolor{gray!20}+0.3\% & \textbf{64.4\%} & \textbf{59.5\%} & \cellcolor{gray!20}-7.6\% & \textbf{65.6\%} & \textbf{59.1\%} & \cellcolor{gray!20}-9.9\% & \textbf{61.7\%} \\
    
\end{tabular}

    \label{tab:33cold_start}
\end{table*}

\subsection{RQ3: Study on Warm and Cold Scenarios}
Most previous studies have focused on warm and cold scenarios for user or/and item \cite{kim2024large, kim2025lostsequencelargelanguage}.
However, interval provides a perspective different from users and items, for evaluating whether interactions occur frequently within a short period.
The interval-based viewpoint can reflect the activity level of each user.

Following the setup of A-LLMRec \cite{kim2024large}, a user or item is categorized as "Warm" if it falls within the top 35\% of interactions and as "Cold" if it falls within the bottom 35\%. 
For intervals, we first compute the average interval between interactions for each user. 
If a user's average interval falls within the top 35\% of interactions, they are classified as "Warm", and those in the bottom 35\% as "Cold". 
After training the model on the full training dataset, we distinguish warm and cold scenarios based on this categorization.


\subsubsection{\textbf{Superiority of IntervalLLM}}
IntervalLLM outperforms all baselines across all warm and cold scenarios, and all perspectives as shown in Table \ref{tab:33cold_start}.
This result highlights that integrating intervals and leveraging their relationships with items enables the LLM to better capture temporal knowledge.
Consequently, our approach not only alleviates the cold-start issue from the interval perspective but also improves performance in user and item cold-start scenarios.

\subsubsection{\textbf{Performance Drop Analysis}}
\emph{Diff.} is used as an additional metric to evaluate the performance drop between warm and cold scenarios.
When comparing methods that incorporate interval information (TiSASRec, LLaMA + Interval, and IntervalLLM) with those that do not, the smaller performance drops in the interval perspective highlight the advantage of leveraging interval information to mitigate the cold-start issue.
Overall, IntervalLLM exhibits the smallest performance drop across both the user and interval perspectives, demonstrating its effectiveness in addressing the cold-start problem.

\subsubsection{\textbf{Performance Drop in the Interval Perspective is more Larger than in User and Item Perspectives}}
A comparative analysis of intervals, users, and items reveals that the interval perspective experiences the most significant performance drop across all methods.
This finding suggests that the interval aspect has been largely overlooked in prior research.
Our study underscores the importance of addressing the cold-start issue specifically from the interval perspective.

\section{Conclusion}
This paper introduces a novel perspective by benchmarking the cold-start scenario from the interval viewpoint, going beyond traditional user and item dimensions, and highlighting a significant performance drop in previous works.
Our proposed IntervalLLM not only encodes dynamic intervals, but also learns the extent to which interval influences item recommendations through interval-infused attention.
The experiments show that empowering LLM by interval yields effectiveness improvements for recommendation models across various benchmarks.
In addition, we analyze the performance from the cold scenario of user, item, and interval, respectively, to identify both well-supported and underperforming user groups, which consistently show the superior performance of IntervalLLM in both warm and cold scenarios.
For future research, we plan to explore other possibilities of integrating interval into LLM and further solving the cold-start interval issue.



\bibliographystyle{ACM-Reference-Format}
\bibliography{reference}


\end{document}